\documentclass [12pt,fleqn]{article}

\usepackage{verbatim}
\usepackage{graphicx}
\usepackage{amssymb}
\usepackage{amsmath}
\usepackage{natbib}                 
\usepackage{enumerate}
\usepackage{tikz}
\usepackage{float}
\usepackage{mathtools}

\newtheorem {lemma} {Lemma}

\newtheorem {corollary} {Corollary}

\newtheorem {proposition} {Proposition}

\newcounter{excount}
\newenvironment{example}
{\refstepcounter{excount}\bigskip\noindent\textbf{Example~\arabic{excount}:}}{$\;\rule{1.5mm}{3mm}$\medskip}

\newcounter{asscount}

\newcommand{\eprf}{$\;\rule{1.5mm}{3mm}$} 

\begin{document}

\title{Optimal Information Acquisition Under Intervention}
\author{Augusto Nieto-Barthaburu\footnote{augusto.nieto@gmail.com. We thank Robert Lieli, Julian Martinez-Iriarte, and Cesar Sosa-Padilla for useful discussions.}\\Department of Economics\\Universidad Nacional de Tucuman}
\maketitle

\thispagestyle{empty}

\begin{abstract}
\begin{footnotesize}
We present a model of a forecaster who must predict the future value of a variable that depends on an exogenous state and on the intervention of a policy-maker. We investigate the incentives of the forecaster to acquire costly private information to use in his forecasting exercise. We show that the policy-making environment plays a crucial role in determining the incentives of the forecaster to acquire information. Key parameters are the expected strength of policy intervention, the precision of the policy-maker's private information, and the precision of public information. We identify conditions, which are plausible in applications, under which the forecaster optimally acquires little or no private information, and instead bases his forecast exclusively on information publicly known at the time the forecast is made. Furthermore we show that, also under plausible conditions, stronger policy intervention and more precise policy-maker's information crowd-out forecaster's information acquisition.

\bigskip



\end{footnotesize}	
\end{abstract}

\section{Introduction}
In this paper we address the problem of optimal information acquisition by a forecaster who must predict the future value of a variable, referred to as the ``outcome,'' that depends both on an exogenous state of the world and on the action of a policy-maker (henceforth PM).\footnote{We will use the pronouns `he' for the forecaster and `she' for the PM.} Our main interest is on how the parameters of the PM's intervention problem affect the forecaster's incentives to acquire costly private information. The following example illustrates the type of environments that are the object of our analysis.

\begin{example}\label{ex: inflation forecasting}
A participant of the Survey of Professional Forecasters (SPF) must make a prediction of future inflation. The forecaster knows that the Federal Reserve's FOMC: $(i)$ will make policy decisions
that will have an impact on future inflation; and $(ii)$ have their own private information on the state of the economy, i.e. on inflationary pressures, that policy decisions will be based on. Therefore, to make a good forecast of inflation the forecaster must predict not only the state of the economy, but also the monetary policy actions that will be taken by the Fed's policy committee. In turn, predicting policy actions requires the forecaster to make inferences about the policy committee's private information, and about how that information translates into policy.
\end{example}


\noindent This paper is an attempt to understand information acquisition incentives in environments like that of Example \ref{ex: inflation forecasting}. While empirically relevant, the analysis of this type of framework has been mostly overlooked in the literature. Our goal is to fill that gap.

We present a formal model a forecaster who must make a prediction of the future value of an economic variable with the objective of minimizing mean squared error (MSE). The realization of the forecasted variable depends on an exogenous state of the world, unknown at the time the forecast is made, and on a policy variable whose value will be chosen by a PM. To make his prediction the forecaster uses private information about the state of the economy that he acquires, in the form of a signal whose precision he can increase at a cost. Hence, when deciding his optimal amount of private information the forecaster trades-off the benefits of a more precise forecast against the costs of a more precise signal. However, since the PM's action has an impact on the forecasted variable, such policy action will typically distort the forecaster's information acquisition tradeoff relative to the no-intervention case. The main focus of our paper is on how the forecaster's information acquisition incentives, and hence the optimal amount of information, change when the policy-making environment changes.

The intuition of why policy distorts information acquisition incentives is the following. Since PM bases her action on her own private information on the state, such action will typically be correlated with the state itself. Therefore, a forward-looking forecaster will anticipate the policy action to either increase or decrease the impact of the state on the outcome, depending on the PM's objectives. More specifically, the PM's action will typically have an impact on variability of the outcome. Given the forecaster's MSE objective, a change in the variability of the outcome will in turn have an effect on the forecaster's loss from predition errors, and therefore on the tradeoff that the forecaster faces when choosing the optimal amount of private information to acquire. To illustrate, consider again the setting of Example \ref{ex: inflation forecasting}.

\setcounter{excount}{0}

\begin{example}
(Cont.) Assume that the FOMC has an inflation target of 2\%, has accurate information about the state of the economy, and has at its disposal policy tools to mostly offset any inflationary pressures that will deviate future inflation from its target. In this case the forecaster can expect that, regardless of the realized state of the economy, the committee's policy actions will lead to future inflation to deviate little from the committee's target, i.e. to be close to 2\%. Therefore, the forecaster has little incentive to acquire private information about the state to make his forecast, given that he expects little variability of the outcome in his forecasting problem.
\end{example}

The framework of Example \ref{ex: inflation forecasting} is surely extreme: typically PMs will have less-than-perfect information about the state of the economy, and also less-than-perfect tools to control outcome realizations. However, the example illustrates the fact that in situations in which economic outcomes depend both on an exogenous state variables and on endogenous policy actions, we can in general expect policy-making to affect the variability of predicted outcomes and, therefore, to also have an impact on forecasters' incentives to acquire private information to make their predictions. Our simple model transparently shows the relationship between the parameters of the PM's intervention problem and the forecasted variable. Key parameters are the expected strength of policy intervention, the precision of the policy-maker's private information, and the precision of public information. Our model sheds light on how changes in these parameters impact the optimal amount of private information for the forecaster.

With the aid of our model we identify conditions under which: $(i)$ the better informed the PM is, the less incentive the forecaster has to acquire information, that is, PM's information crowds-out forecaster's information; and $(ii)$ the stronger the policy action expected by the forecaster, the less informed the forecaster becomes. Indeed, we show that under some conditions the forecaster will have very little incentive to get privately informed; and furthermore, we argue that such conditions are plausible to hold in important applications.

More specifically, we show that in contexts in which the PM's action is ``preventive,'' and she is well informed about the decision-relevant state of the world, the forecaster will optimally acquire little or no information for making his prediction.\footnote{The PM's action is preventive when her aim is to counteract the effect of the state on the outcome, as in Example \ref{ex: inflation forecasting}.} In those cases private forecasts will be based mostly on information that is publicly known at the time of the forecasting exercise. This result is intuitive: if the PM is well informed and follows a preventive policy rule, the forecaster can expect the value of the outcome to be ``smoothed out'' by the PM's action. Given that, the prediction errors that the forecaster expects to make are small, and therefore his marginal benefit from acquiring private information is low.

We also show that when the PM's action is preventive the effect of better public information is generally ambiguous. The reason is that better public information has a dual impact on incentives. First, it substitutes for private information. Second, and more subtlety, it lowers the incentives \emph{of the PM} to act following her own private signal. That is, it leads the PM to follow the public signal more (and his private signal less). When the PM's action is preventive such weaker response of the PM to her signal leads to an increase in the variability of the outcome. This indirect effect increases the forecaster's marginal benefit of private information, making the sign of the total effect ambiguous.

Taken together, our findings cast doubts on the informativeness and social value of private forecasts in environments in which policy intervention is consequential. Indeed, our results are consistent with empirical findings in that direction. In a study that analyzes forecasts of US GDP growth and inflation from a panel of professional forecasters, \cite{patton:2010aa} find that individual forecasts differ very little at short horizons, and some more at long horizons.
\cite{patton:2010aa} interpret their findings as indicating that: $(i)$ individual forecasts are based on essentially no private information --- i.e. that all forecasters use the same set of publicly available variables in their forecasts; and $(ii)$ dispersion of forecasts at long horizons is based on differences in priors and/or models, instead of on differing private signals. Notably, the findings in \emph{ibid.} are relative to predictions of GDP growth and inflation, two variables that are arguably affected by policy (inflation perhaps more than GDP). Our model shows that the findings of \emph{ibid.} are to be expected in such a context, i.e. that private forecasters will typically have little incentive to collect proprietary data when outcomes are expected to have limited volatility due to policy.


Finally, in an extension of our analysis we consider the case in which the PM publicly discloses her private signal. We show that such PM transparency has a dual role on the forecaster's information acquisition incentives. First, it provides the forecaster with a free signal of the state, hence making the state less uncertain and therefore lowering incentives for information gathering. Second, it makes the PM's action perfectly predictable for the forecaster, who can now account for it exactly in the forecast. This second effect lowers the volatility of the ``residual'' (i.e. policy-adjusted) outcome. We identify two cases: $(i)$ when the policy action \emph{reinforces} the effect of the state on the outcome, transparency eliminates that perception of reinforcement, and therefore further diminishes incentives for information acquisition; $(ii)$ when the policy action \emph{mitigates} the effect of the state on the outcome, transparency eliminates such mitigating effect, and hence it increases the incentives for information acquisition. In sum, when the policy action is reinforcing both effects work in the same direction, and hence transparency unequivocally lowers incentives for acquiring information; on the other hand, when policy actions are preventive the effects work on opposite directions, and the effect of transparency on information acquisition is ambiguous.

We point out that while for concreteness here we focus on the problem of a forecaster, the analysis in this paper applies more generally. Indeed, our model is appropriate for any information acquisition problem in which: $(i)$ there is an agent who must make a decision whose payoff depends on a stochastic outcome and on the agent's own action; and $(ii)$ the stochastic payoff-relevant outcome depends on an exogenous state and on the policy action of another agent (the PM in our paper). An example of an alternative application that also fits our setup is that of a bond investor who must decide to acquire private information to make an investment. Typically the payoff of a bond investor will depend on future interest rates, which in turn depend on the future state of the economy and on monetary policy. All our results extend naturally to such settings.

\subsection{Related literature}
The literature on optimal information acquisition is large. Notably, recent contributions on Rational Inattention pioneered by \cite{sims:2003aa} have provided a flexible framework for analyzing costly information acquisition in various settings. The Gaussian-quadratic structure on which our model is built is a special case of that framework. In this paper we augment that Gaussian-quadratic model with an intervener (the PM) who makes a policy decision that has an impact on the payoff-relevant outcome.
To the best of our knowledge we are the first to embed a policy action affecting the payoff-relevant outcome to study the effects of policy-making on information acquisition incentives.

A recent literature has looked at private information acquisition in coordination settings; see, e.g., \cite{colombo:2014aa}, \cite{hellwig:2009aa}, and \cite{mackowiak:2009aa}. These papers point out that when economic agents have coordination (or mis-coordination)  motives, they may want to acquire private information to predict what others will do, in addition to predicting an exogenous payoff-relevant state. A common finding in this literature is that better public information crowds out private information. None of these papers feature a PM who has an effect on the payoff-relevant outcome. In contrast to their results, in our model better public information may either increase or decrease the optimal amount of forecaster's private information, since public information has both a direct and an indirect effect on the incentives of the forecaster, as explained above.


A related literature has looked at the effects of public information precision on information aggregation. In particular, \cite{morris:2002aa, morris:2005aa} show that in beauty-contest settings more precise public information helps agents coordinate better, leading them to follow their (exogenous) private information  less in their decisions, and therefore to less information aggregation by prices. These papers don't analyze how private information acquisition incentives are affected by better public information, which is one of the questions that we tackle.

The rest of the paper proceeds as follows. In Section \ref{sec: model} we present our formal model.
In Section \ref{sec: analysis} we analyze the model, and in particular we derive the optimal information acquisition strategy for the forecaster. In Section \ref{sec: comp statics} we derive comparative statics results of optimal forecaster's information precision with respect to the various parameters of the PM's problem. Section \ref{sec: transparency} presents an extension in which the PM reveals her signal publicly, and analyze whether PM transparency encourages or inhibits private information acquisition. Section \ref{sec: conclusion} concludes. All proofs are contained in the Appendix.

\section{Model}\label{sec: model}
We consider the problem of a forecaster who must predict the future value of a variable $y$ (the ``outcome''), which is a function of the state of the economy and of a policy variable which will be decided by a PM. Concretely, we assume that
\begin{align*}
y=\theta+a+\epsilon,
\end{align*}
where $\theta$ is the unknown state of the economy, $a$ is the policy action of the PM, and $\epsilon$ is a zero-mean random variable, which is independent of $\theta$ and has variance $\tau^2$.  We assume that $\theta\sim N(\bar{\theta},\frac{1}{t})$ -- that is, $\bar{\theta}$ is the mean and $t$ is the precision (the inverse of the variance) of the prior distribution of the state. To simplify computations we normalize the prior mean to zero, i.e. we assume $\bar{\theta}=0$.

Before making her decision, the PM observes a private signal
\begin{align*}
s_P=\theta+\eta,
\end{align*}
where $\eta\sim N(0,\frac{1}{h})$, and is independent of $\theta$ and $\epsilon$. The precision of the PM's signal is assumed exogenous.\footnote{The justification for this assumption is that since in our model the PM does not use the forecaster's prediction to inform her decision, the optimal value of her signal's precision does not depend on the forecaster's signal precision (to be introduced below). Hence, we do not need to explicitly model the PM's information acquisition problem, and therefore we can take her precision as exogenously given. The exogenous PM's precision $h$ can be interpreted as that resulting from a prior (unmodelled) optimization problem.}
Using the information in her signal, the PM makes her policy decision following the rule
\begin{align}\label{eq: PM's action}
a^*=-x \cdot E(\theta|s_P).
\end{align}
In equation (\ref{eq: PM's action}) the random variable $x$ measures the \emph{strength of intervention}. We assume that the value of $x$ is private information of the PM.\footnote{For example, it is typically the case that outsiders (and even insiders) to government policy-making bodies have some degree of uncertainty about upcoming policy decisions. As an illustration, consider monetary policy decisions: there is seldom consensus on the future path of interest rates that the Fed's policy committee will adopt, even among members of such committee. Therefore, it is natural to assume that the forecaster has some degree of uncertainty about upcoming policy decisions that will have an impact on the accuracy of his forecasts.}  The forecaster's beliefs about $x$ are given by a probability distribution with support in a subset of $[-\bar{x},\bar{x}]$ for some $\bar{x}>0$. Furthermore, we assume that $x$ is independent of $\theta$, $\epsilon$ and $\eta$, and we denote its mean and variance by $\mu$ and $\sigma^2$, respectively.\footnote{The independence between $x$ and other variables is assumed mainly for analytical tractability. In applications it is conceivable, however, that the strength of policy $x$ may be correlated with the state of the economy $\theta$. For example, we would typically expect stronger monetary policy actions in the time of a financial crisis or a period of high inflationary pressures.} Given the PM's optimal action, the outcome to be predicted by the forecaster is
\begin{align*}
y=\theta-x E(\theta|s_P)+\epsilon.
\end{align*}

While we are specifying the PM's action exogenously, a decision rule of the form (\ref{eq: PM's action}) can be easily micro-funded. Consider, for example, the case of a PM who has the objective of steering the realization of the random variable $y$ toward a target value $y^T$ (implicitly normalized to zero), and must pay a cost of intervention which is convex in the policy variable $a$. More concretely, assume that the PM has a policy loss function which is the sum of the square of the distance of $y$ from a target value (normalized to zero) plus a quadratic cost of intervention. Such model of PM's behavior is used in \cite{lieli:2026aa}, which derive a policy rule of the form (\ref{eq: PM's action}) as optimal for the PM in that context. In that model of PM behavior the value of $x$ depends on the cost of intervention relative to the cost of suffering deviations of $y$ from its target value. When $x$ is positive, the PM's optimal action is \emph{preventive}, in the sense that it is aimed at counteracting the effect of the state on the outcome. Such is the case, for example, in the model of \cite{lieli:2026aa}. On the other hand, a PM who has a negative optimal value of $x$ wishes to \emph{reinforce} the effect of the state on the outcome.

We now switch to the forecaster. To make his prediction of $y$ he uses the information contained in a signal he collects, given by
\begin{align*}
s=\theta+\delta,
\end{align*}
where $\delta\sim N(0,\frac{1}{k}$). For simplicity we assume that $\delta$ is independent of all other random variables above, i.e., of $\theta$, $\epsilon$, $\eta$, and $x$. While we assume that the forecaster's information structure is exogenously normal, the forecaster can choose the precision $k$ of his signal at a cost $C(k)$, which is taken to be increasing and convex.\footnote{For example, the forecaster can base his prediction on a smaller or larger private sample, and collecting a larger sample may be more costly.} We note that we allow for the cost function to be linear, i.e. strict convexity is not assumed.

As is standard in the forecasting literature, we assume that the forecaster's loss function is quadratic, and therefore that the objective of his forecasting exercise is to minimize the mean-squared-error (MSE) of his forecast. In addition, he pays the cost of acquiring information. Hence, the forecaster's ex-ante problem is
\begin{align}\label{eq: forecaster's problem}
\min_{k\ge 0} E\{\min_{f}E[(y-f)^2|s]+C(k)\}=\min_{k\ge 0} E\{\min_{f}E[(\theta+a^*+\epsilon-f)^2|s]+C(k)\}.
\end{align}
More explicitly, after having chosen precision level $k$ for his signal, the forecaster's interim objective is to minimize MSE conditional on the realized value of his signal. That forecasting problem is represented by the minimization problem in the first term inside the curly brackets in (\ref{eq: forecaster's problem}) -- the inner expectation is with respect to the state $\theta$ conditional on signal realization $s$. In turn, the forecaster's ex-ante problem is to choose the optimal precision $k$ for his signal, given that his signal will then be used optimally in the interim forecasting problem. The choice of the optimal precision is represented by the outer minimization problem in (\ref{eq: forecaster's problem}), in which the forecaster trades-off the gains of higher accuracy of a more precise signal in predicting $y$ against the costs of acquiring a signal with higher precision --- the outer expectation is with respect to the distribution of the signal $s$.

Figure \ref{fig: timeline} shows the timeline of our model. The nodes in blue represent the moments of choice of the endogenous variables by the forecaster, while the moments in black represent the realization of the exogenous variables.

\bigskip

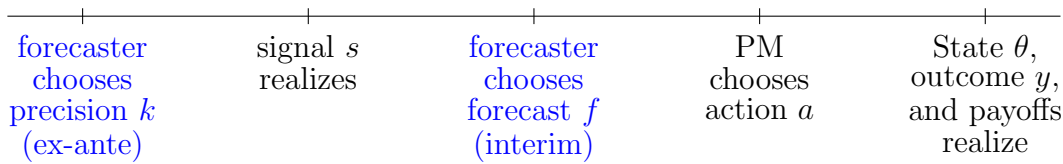
\begin{center}
\begin{tikzpicture}\label{fig: timeline}
    
    \draw (0,0) -- (14,0);

    \foreach \x in {1,4,7,10,13}
    \draw (\x cm,3pt) -- (\x cm,-3pt);

   \draw (1,0) node[below=3pt] {\textcolor{blue}{forecaster}} node[below=15pt] {\textcolor{blue}{chooses}} node[below=27pt] {\textcolor{blue}{precision $k$}} node[below=39pt] {\textcolor{blue}{(ex-ante)}};
   \draw (4,0) node[below=3pt] {signal $s$} node[below=15pt] {realizes};
   \draw (7,0) node[below=3pt] {\textcolor{blue}{forecaster}} node[below=15pt] {\textcolor{blue}{chooses}} node[below=27pt] {\textcolor{blue}{forecast $f$}} node[below=39pt] {\textcolor{blue}{(interim)}};
   \draw (10,0) node[below=3pt] {PM} node[below=15pt] {chooses} node[below=27pt] {action $a$};
   \draw (13,0) node[below=3pt] {State $\theta$,} node[below=15pt] {outcome $y$,} node[below=27pt] {and payoff} node[below=39pt] {realize};
  \end{tikzpicture}
\end{center}


\paragraph{Interpretation of the prior} In our model, both the PM and the forecaster share a common prior on $\theta$, and in addition they receive private signals $s_P$ and $s$ respectively. The precision of the PM's private signal is exogenous, and the precision of the forecaster's signal, which is our main interest, is endogenously determined in the model. In that context, it is natural to interpret the prior as the shared beliefs of the PM and the forecaster \emph{before} receiving their respective private signals, i.e. as the beliefs resulting from (unmodeled) publicly known information. With that interpretation, the prior mean $\bar{\theta}$ (which is normalized to zero) serves as a public signal of the state, and the prior precision is a measure of precision of public information. That interpretation will allow us to inquire about the effect of better (or worse) public information on the forecaster's information acquisition incentives.

In the next section we analyze the problem of the forecaster, and derive both his optimal interim forecast and his optimal ex-ante choice of precision. An immediate result, stated in Proposition \ref{proposition: precisions comparison intervention no intervention}, is that the optimal precision of the forecaster will typically be lower under intervention that when such intervention is absent. Indeed, we show by example that in plausible scenarios the incentives of the forecaster to acquire private information may be completely crowded out, and in such cases forecasts will be based solely on public information.

\section{Analysis}\label{sec: analysis}
To solve the forecaster's problem we proceed by backward induction. That is, we start from the interim problem of choosing the optimal forecast $f^*$ having received signal $s$ with previously chosen precision level $k$, and then we move to the problem of finding the optimal precision $k^*$ given the results of the interim forecasting problem.

\subsection{Interim choice of the optimal forecast}

In the interim period the forecaster has already observed the realized signal $s$ with previously chosen precision $k$. Therefore, at that point he solves
\begin{align}\label{eq: interim forecaster's problem}
\min_{f}E[(y-f)^2|s] = & \min_{f}E[(\theta+a^*+\epsilon-f)^2|s] \nonumber\\
= & \min_{f}E[(\theta+a^*-f)^2|s]+\tau^2.
\end{align}
Given the Gaussian information structure, we can explicitly compute the PM's action $a^*$. From equation (\ref{eq: PM's action}), it takes the form
\begin{align}\label{eq: PM's action 2}
a^*= &-x E(\theta|s_P) \nonumber\\
=& -x\frac{h}{t+h}s_P \nonumber\\
=& -x H s_P,
\end{align}
where we denote $H \equiv \frac{h}{t+h}$. In (\ref{eq: PM's action 2}) we have used the standard form for the conditional expectation for a normal information structure, plus the fact that the prior mean is normalized to zero. Parameter $H$ is the weight that the PM puts on her signal to compute $E(\theta|s_P)$; the weight on the prior mean is $(1-H)$, but since the prior mean is normalized to zero, it does not show in the expression for the PM's action (\ref{eq: PM's action 2}).

Substituting the PM's action (\ref{eq: PM's action 2}) into the forecaster's interim loss function (\ref{eq: interim forecaster's problem}), the forecasting problem becomes
\begin{align*}
\min_{f} E[(\theta-x H s_P-f)^2|s]+\tau^2.
\end{align*}
Using the fact that that $s_P=\theta+\eta$, we can further expand the forecasting problem as
\begin{align}\label{eq: interim forecaster's problem 2}
& \min_{f} E[(\theta-x H \theta-x H \eta-f)^2|s]+\tau^2 \nonumber \\
& =\min_{f} E\{[(1-x H ) \theta -x H \eta-f]^2|s\}+\tau^2  \nonumber \\
& = \min_{f} E\{[(1-x H ) \theta-f]^2 | s\} + H^2 E(x^2) E(\eta^2)+\tau^2,
\end{align}
where the last equality uses the fact that $\theta$, $x$, $\eta$ are mutually independent.

The last two terms in equation (\ref{eq: interim forecaster's problem 2}) do not depend on $f$, and hence are irrelevant for the choice of the optimal forecast. Given that, the form of the optimal forecast in the next lemma follows from standard results for quadratic loss functions, and the form of the optimal interim loss follows from direct calculation after substituting the optimal forecast into the interim loss. The proof, as all others in the paper, is relegated to the Appendix.

\begin{lemma}\label{lemma: optimal forecast and interim loss}
\begin{enumerate}
\item The optimal forecast is
\begin{align*}
f^*= & E[(1-x H )\theta|s].
\end{align*}

\item The optimal interim loss for the forecaster is
\begin{align}\label{eq: optimal interim loss}
Var [(1-x H )\theta |s ]+ H^2 E(x^2) E(\eta^2) + \tau^2.
\end{align}
\end{enumerate}
\end{lemma}
The expression for the optimal interim loss (\ref{eq: optimal interim loss}) shows clearly the three different sources of loss for the forecaster in the interim period. The first term captures the variability in the fraction of the state that is uncorrected by the PM, conditional on the forecaster's information.\footnote{Recall that the PM corrects a fraction $x \cdot H$ of the state with her action, and therefore only the remaining uncorrected fraction of the state goes into the outcome $y$. We emphasize that ``uncorrected fraction'' is to be read generally, since if $(1-x H )>1$ the PM's action \emph{adds} variability to the outcome. This will be the case if $x<0$.} This variance is the optimal forecasting loss, given that the optimal forecast is the conditional expectation; importantly, this variance can be reduced by acquiring a more precise signal in the ex-ante period. The second term captures the loss caused by the fact that since the correction of the state by the PM is imperfect, given her imperfect signal, unforecastable errors in the PM's signal translate into errors in policy, that in turn translate into variability in the outcome that is unpredictable for the forecaster. Finally, the last term captures the loss due to errors in predicting the outcome $y$ caused by the unforecastable shock $\epsilon$.

The last two terms in the optimal interim loss (\ref{eq: optimal interim loss}) do not depend on the forecaster's signal precision $k$, and are therefore irrelevant for the choice of optimal precision. Hence, when considering the choice of optimal precision we can consider the forecaster's optimal interim loss to be just
\begin{align}\label{eq: optimal interim loss 2}
Var [(1-x H )\theta |s ].
\end{align}

\subsection{Ex-ante choice of optimal precision}
We now turn to the ex-ante problem of choosing the optimal precision. In the ex-ante period the forecaster chooses optimal precision $k$ to minimize the expected value of (\ref{eq: optimal interim loss 2}) plus the cost of precision $C(k)$. That is, the ex-ante problem of the forecaster is
\begin{align}\label{eq: forecaster's ex-ante problem}
\min_{k\ge 0} E\{Var [(1-x H )\theta |s ]\}+C(k),
\end{align}
where the expectation is with respect to signal $s$. From the Law of Total Variance, we can write the first term in (\ref{eq: forecaster's ex-ante problem}) as
\begin{align}\label{eq: forecaster's ex-ante problem 2}
E\{Var [(1-x H )\theta |s ]\}=Var [(1-x H )\theta ] - Var\{E[(1-x H)\theta |s]\}.
\end{align}
Expression (\ref{eq: forecaster's ex-ante problem 2}) has an intuitive interpretation: the ex-ante expected squared forecast errors are equal to the total variance of the predicted variable $(1-x H )\theta$ minus the part of that variance that is explained by the predictor $E[(1-x H)\theta |s]$. This second term can be increased by the acquisition of a more precise signal, at the expense of a higher cost. That is the fundamental tradeoff that the forecaster faces in the ex-ante period. Direct calculation of the terms in (\ref{eq: forecaster's ex-ante problem 2}) leads to the following lemma.
\begin{lemma}\label{lemma: ex-ante loss}
The forecaster's ex-ante problem (\ref{eq: forecaster's ex-ante problem}) can be written as
\begin{eqnarray}\label{eq: ex ante problem 3}
\min_{k\ge 0}  \left\{(1-\mu H)^2 Var(\theta|s)  + C(k)\right\}=\min_{k\ge 0}  \left\{(1-\mu H)^2 \left(\frac{1}{t+k} \right) + C(k)\right\}.
\end{eqnarray}
\end{lemma}

\subsection{Forecaster's problem with and without intervention}
At this point it is instructive to think on the special case in which there is no intervention by the PM, which is the standard setting in the forecasting literature. That case corresponds to $(\mu=0, \sigma^2=0)$ in our model. From (\ref{eq: ex ante problem 3}) we see that in that situation the forecaster's ex-ante problem becomes 
\begin{eqnarray}\label{eq: forecaster's ex ante problem no intervention}
\min_{k\ge 0} \left\{Var(\theta|s)+C(k)\right\} = \min_{k\ge 0} \left\{\frac{1}{t+k}+C(k)\right\}.
\end{eqnarray}
The intuition of the no-intervention case is simple. Absent intervention, the problem of the forecaster reduces to that of predicting the state $\theta$. Given his MSE loss, his best interim prediction is the conditional expectation of the state given his signal, and therefore his optimal interim loss is the conditional variance of the state given his signal, which is given by the first term in (\ref{eq: forecaster's ex ante problem no intervention}). (With a Gaussian information structure the conditional variance of the state is a constant.) Hence, his ex-ante problem in the no-intervention case is to choose his signal's precision $k$ trading-off expected losses from predicting the state with the cost of acquiring precision.

The intuition of the no-intervention case helps to understand that of PM's intervention. In the more complex case with PM intervention the forecaster also trades-off expected forecast errors of the state with the cost of precision, as in the simpler no-intervention setting. However, PM's intervention changes the variability of the outcome through her action. In particular, the variance of the outcome due to uncertainty about the state is strictly reduced by the PM's intervention if and only if
\begin{eqnarray}\label{eq: PM correction factor}
(1- \mu H)^2=\left(1- \mu \frac{ h}{t+h}\right)^2 < 1,
\end{eqnarray}
and it is strictly increased if and only if the strict inequality is reversed. When condition (\ref{eq: PM correction factor}) holds, the forecaster expects the outcome $y$ to be less variable than in the no-intervention case, because the PM corrects part of the state with her action, and therefore that portion of the state does not translate into variability of the outcome $y$. Hence, the forecaster expects lower forecast errors for any given level of precision. On the other hand, when condition (\ref{eq: PM correction factor}) is reversed, PM's intervention reinforces the effect of the state on the outcome, and therefore the forecaster expects higher errors for any given level of the signal's precision.


The discussion in the preceding two paragraphs leads to the following result.
\begin{proposition}\label{proposition: precisions comparison intervention no intervention}
In the forecaster's problem with intervention:
\begin{enumerate}
\item The optimal precision $k^*$ is weakly lower than in the no-intervention case if
\begin{eqnarray}\label{eq: condition 1 lemma 1}
\left(1- \mu \frac{ h}{t+h}\right)^2 < 1      \iff       0<\mu <2 \frac{t+ h}{h}.
\end{eqnarray}

\item The optimal precision $k^*$ is weakly higher than in the no-intervention case if
\begin{eqnarray*}\label{eq: condition 2 emma 1}
\left(1- \mu \frac{ h}{t+h}\right)^2 > 1     \iff      \mu\in [-\bar{x},0) \cup \left(2\frac{t+ h}{h},\bar{x}\right].
\end{eqnarray*}
\end{enumerate}
\end{proposition}
Intuitively, the benefit from acquiring higher precision in the ex-ante period comes from lowering the conditional variance of the state in the forecasting exercise of the interim period. When there is intervention, the weight of the conditional variance relative to the cost of precision in the ex-ante forecaster's loss is lower (higher) whenever condition (\ref{eq: condition 1 lemma 1}) holds (is reversed), and therefore the incentives of the forecaster to acquire precision are lower (higher) than in the no-intervention case. We believe that condition (\ref{eq: condition 1 lemma 1}) is plausible in many applications; it will be reasonably expected to hold, for example, when the PM takes corrective actions that counteract the effect of the state on the outcome, as long as she doesn't overreact substantially to her information.\footnote{
One application in which condition (\ref{eq: condition 1 lemma 1}) seems particularly reasonable is inflation forecasting, in which monetary policy is typically thought to have a preventive nature, trying to correct deviations of expected inflation from a set target.}

An implication of Proposition \ref{proposition: precisions comparison intervention no intervention} is that in environments in which the PM's actions are preventive and the precision of public and PM's information is high, the incentives of the forecaster to acquire private information will typically be low. Indeed, in the problem with intervention it is not difficult to generate corner solutions in which $k^*=0$, while for the same parameter values in the problem with no intervention $k^*>0$. The following example illustrates.

\begin{example}\label{ex: corner solution}
Assume that $t=1$, $h=1$, and $\mu=1$. Furthermore, assume that the cost of acquiring precision for the forecaster is linear, $C(k)=\frac{k}{2}$. Then the forecaster's problem with no intervention is
\begin{eqnarray*}\label{eq: forecaster's ex ante problem no intervention example}
\min_{k\ge 0} \left\{ \frac{1}{1+k}+\frac{k}{2}\right\}.
\end{eqnarray*}
which has solution $k^*=\sqrt{2}-1>0$. On the other hand, the forecaster's problem with intervention is
\begin{eqnarray*}\label{eq: forecaster's ex ante problem no intervention example}
\min_{k\ge 0} \left\{\frac{1}{4} \frac{1}{1+k}+\frac{k}{2}\right\}.
\end{eqnarray*}
which has solution $k^*=0$.
\end{example}

Example \ref{ex: corner solution} shows that intervention can have a significant impact on the incentives of the forecaster to acquire information. We see in the example that for some parameter values the forecaster optimally acquires no private information under intervention, and instead makes his forecast based solely on public information (i.e. on the prior), while he would acquire some private information if intervention were absent. Hence, in the example PM's intervention completely eliminates information acquisition incentives for the forecaster.

The observation that in some situations public and PM's information may completely crowd-out private information is consistent with previous findings in the forecasting literature. For instance, \cite{patton:2010aa} show that forecasters in their sample base their inflation forecasts on essentially no private information. We observe that in the setting of \cite{patton:2010aa}: $(i)$ public information about inflationary pressures, in the form of government statistics and other publicly available market data, is typically quite good --- in the notation of our model, $t$ is high; $(ii)$ the Fed (the PM) typically takes strong corrective actions to fight deviations of inflation from her target --- in the notation of our model, $\mu$ is close to 1; and $(iii)$ the Fed expends significant effort to acquire additional information to base monetary policy on, as evidenced by the substantial resources that are devoted to making Green Book and other internal predictions --- in the notation of our model, $h$ is high. In the light of Proposition \ref{proposition: precisions comparison intervention no intervention}, it is not surprising then that the incentives of private forecasters to acquire proprietary data in that environment will be low, and therefore that various private forecasts will mostly be based on the same publicly available information.


In the next section we derive comparative statics results for the forecaster's optimal precision. The goal is to understand the effect of each of the parameters of the PM's intervention problem on the incentives of the forecaster to acquire private information.

\section{Comparative statics}\label{sec: comp statics}

In this section we conduct comparative statics exercises with respect to the various parameters in our model. In particular, we will analyze how the optimal precision of the forecaster depends on three key parameters: $(i)$ the precision of the PM's information $h$; $(ii)$ the average strength of PM's intervention $\mu$; and $(iii)$ the precision of public information $t$.

Recall that the forecaster's ex-ante problem is
\begin{eqnarray*}
\min_{k\ge 0}  \left\{(1-\mu H)^2 \left(\frac{1}{t+k} \right) + C(k)\right\} = \min_{k\ge 0}  \left\{\left(1-\mu \frac{h}{t+h}\right)^2 \left(\frac{1}{t+k} \right) + C(k)\right\}.
\end{eqnarray*}
In the analysis below it will be insightful to write the forecaster's choice of optimal precision as a maximization problem:
\begin{eqnarray*}
\max_{k\ge 0} \left\{-\left(1-\mu \frac{h}{t+h}\right)^2 \left(\frac{1}{t+k} \right) - C(k)\right\}=\max_{k\ge 0} \left\{g(k; h,t,\mu)-C(k)\right\},
\end{eqnarray*}
where
\begin{align*}
g(k; h,t,\mu)=-\left(1-\mu \frac{h}{t+h}\right)^2  \left(\frac{1}{t+k}\right).
\end{align*}
The function $g(k; h,t,\mu)$ is the benefit of information precision, which is negative because it is the negative of the forecasting loss. It is easily checked that $g(k; h,t,\mu)$ is increasing in the precision $k$. The partial derivative
\begin{align}\label{eq: marginal benefit of precision}
\frac{\partial g(k; h,t,\mu)}{\partial k}= \left(1-\mu \frac{h}{t+h}\right)^2  \left(\frac{1}{t+k}\right)^2 \ge 0
\end{align}
is the marginal benefit of acquiring information.

It will be seen that the comparative statics for the optimal precision $k^*$ depend on the values of the parameters of the model. To interpret these conditions, recall the form of the PM's action (\ref{eq: PM's action})
\begin{eqnarray} \label{eq: PM's action 3}
a^*=-x \cdot E(\theta|s_P),
\end{eqnarray}
and that $\mu=E(x)$. Consider the following three cases:
\begin{enumerate}
\item $\mu<0$: in this case the PM takes, on average, a \emph{reinforcing} action. As equation (\ref{eq: PM's action 3}) shows, when $\mu$ (the mean of $x$) is negative, on average the PM's action goes in the same direction as the expected value of the state, and therefore it reinforces the expected effect of the state on the outcome $y$. This is the case, for instance, of a PM whose preferences are strictly increasing in the outcome $y$.

\item $0<\mu<\frac{t+h}{h}$: in this case the PM takes \emph{preventive} action. As equation (\ref{eq: PM's action 3}) shows, when $\mu$ is positive, on average the PM's action goes in opposite direction than the expected state, and therefore it counteracts the expected effect of the state on the outcome $y$. This is the case, for example, of a PM who has a target value for the outcome $y$, and uses her action to steer the outcome towards the target (which in our model is normalized to zero). However, the condition that $\mu<\frac{t+h}{h}$ also means that this PM does not substantially overreact to the expected value of the state.

\item $\mu>\frac{t+h}{h}>1$: in this case the PM takes preventive action, but she overreacts to the expected state, i.e. on average the corrective action is strictly greater than what is needed (in expectation) to bring down the outcome to her target value. This overreaction can happen, for example, when the PM has other incentives beyond that of controlling the outcome $y$. For instance, the PM may be inclined to take strong corrective action for political reasons, or to ``send a message'' to establish a reputation.
\end{enumerate}

In the following proposition we state the comparative statics for the forecaster's marginal benefit of information (\ref{eq: marginal benefit of precision}) in the three cases discussed above.
\begin{proposition}\label{prop: CS on marginal benefit}
\begin{enumerate}
\item If $\mu<0$, then the marginal benefit of forecaster's precision (\ref{eq: marginal benefit of precision}) is strictly decreasing in $\mu$, strictly increasing in $h$, and strictly decreasing in $t$.
\item If $0<\mu<\frac{t+h}{h}$, then the marginal benefit of forecaster's precision (\ref{eq: marginal benefit of precision}) is strictly decreasing in $\mu$, strictly decreasing in $h$, and it may be increasing or decreasing in $t$.
\item If $\mu>\frac{t+h}{h}$, then the marginal benefit of forecaster's precision (\ref{eq: marginal benefit of precision}) is strictly increasing in $\mu$, strictly increasing in $h$, and strictly decreasing in $t$.
\end{enumerate}
\end{proposition}
The results in Proposition \ref{prop: CS on marginal benefit} imply the following comparative statics on the optimal precision.
\begin{corollary}\label{cor: CS on k}
\begin{enumerate}
\item If $\mu<0$, then the optimal precision $k^*$ is weakly decreasing in $\mu$, weakly increasing in $h$, and weakly decreasing in $t$.
\item If $0<\mu<\frac{t+h}{h}$, then the optimal precision $k^*$ is weakly decreasing in $\mu$, weakly decreasing in $h$, and it may increase or decrease with $t$.
\item If $\mu>\frac{t+h}{h}$, then the optimal precision $k^*$ is weakly increasing in $\mu$, weakly increasing in $h$, and weakly decreasing in $t$.
\end{enumerate}
\end{corollary}
The results in Corollary \ref{cor: CS on k} follow directly from Proposition \ref{prop: CS on marginal benefit} and standard monotone comparative statics results (\cite{Milgrom:1994aa}, Theorems 5 and 6). All comparative statics in Corollary \ref{cor: CS on k} are strict for interior solutions, that is, for parameter values such that $k^*>0$ (\cite{vanzandt:2002aa}). Figure 1 summarizes these results.

\bigskip

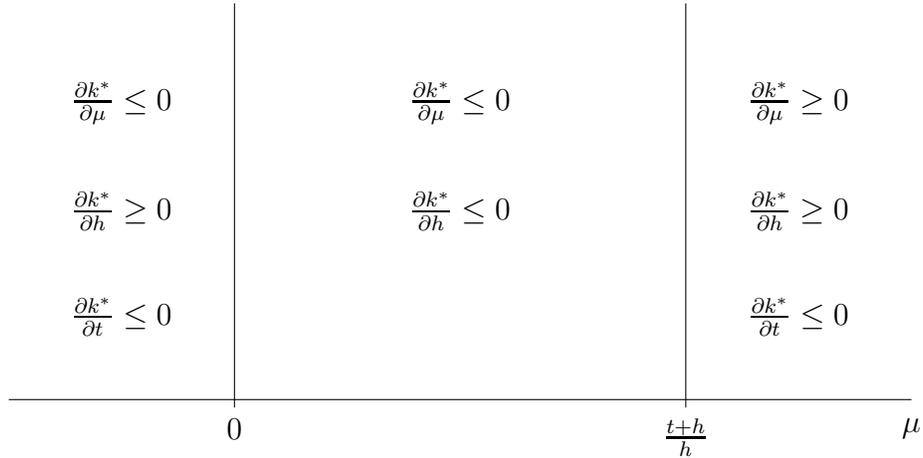
\begin{figure}[h]
\centering
\begin{tikzpicture}
\draw (3,0) -- (15,0);

\foreach \x in {6,12}
\draw (\x cm,150pt) -- (\x cm,-3pt);

\draw (6,0) node[below=3pt] {$0$};
\draw (12,0) node[below=3pt] {$\frac{t+h}{h}$};
\draw (15,0) node[below=3pt] {$\mu$};

\draw (4.5,0) node[above=100pt] {$\frac{\partial k^*}{\partial \mu}\le 0$};
\draw (4.5,0) node[above=60pt] {$\frac{\partial k^*}{\partial h}\ge 0$};
\draw (4.5,0) node[above=20pt] {$\frac{\partial k^*}{\partial t}\le 0$};

\draw (9,0) node[above=100pt] {$\frac{\partial k^*}{\partial \mu}\le 0$};
\draw (9,0) node[above=60pt] {$\frac{\partial k^*}{\partial h}\le 0$};

\draw (13.5,0) node[above=100pt] {$\frac{\partial k^*}{\partial \mu}\ge 0$};
\draw (13.5,0) node[above=60pt] {$\frac{\partial k^*}{\partial h}\ge 0$};
\draw (13.5,0) node[above=20pt] {$\frac{\partial k^*}{\partial t}\le 0$};

\end{tikzpicture}

\caption{Comparative statics for optimal precision $k^*$ (inequalities are strict for $k^*>0$)}
\end{figure}

To gain intuition on the comparative statics results of Corollary \ref{cor: CS on k} we focus on case 2, i.e., the case in which
\begin{eqnarray}\label{eq: CS condition}
0<\mu<\frac{t+h}{h}.
\end{eqnarray}
When condition (\ref{eq: CS condition}) holds, the action of the PM is preventive ($\mu>0$), but the PM does not overreact to the expected state. In this case Corollary \ref{cor: CS on k} shows that the optimal precision is decreasing in both $\mu$ and $h$. The reason is that an increase in $\mu$ and/or $h$ leads to stronger intervention by the PM. In turn, when condition (\ref{eq: CS condition}) holds, stronger intervention makes the outcome \emph{less volatile}. Indeed, under condition (\ref{eq: CS condition}) action is preventive, and therefore the PM subtracts some of the variability of the state from the outcome with her action. Since the outcome to be predicted then becomes less volatile, the ex-ante marginal gain for the forecaster of acquiring private information about the state decreases, and therefore the forecaster optimally acquires less information.

On the other hand, the effect of public information precision $t$ on the optimal forecaster's precision is ambiguous. First, there is a direct effect by which better public information reduces the marginal benefit of information. This direct effect is captured by the fact that the second factor in equation (\ref{eq: marginal benefit of precision}) is decreasing in $t$. Second, there is an indirect effect by which more precise public information leads the PM to weaker preventive action: under condition (\ref{eq: CS condition}) the first factor in (\ref{eq: marginal benefit of precision}) is increasing in $t$.
Hence, the effect on $k^*$ of an increase in $t$ is ambiguous. However, for values of $\mu$ close to 0 or to $\frac{t+h}{h}$ the direct effect dominates, and therefore for those values a small increase in the precision of public information $t$ leads the forecaster to acquire less private information, i.e. $k^*$ decreases with a small increase in $t$.

We note that the optimal forecaster's precision $k^*$ is independent of policy uncertainty $\sigma^2$ (the variance of $x$). Indeed, while $\sigma^2$  does show in the ex-ante expected loss of the forecaster (see expression (\ref{eq: forecaster's ex-ante problem app 3}) in the Appendix), it does so in terms that do not interact with $k$, and therefore it is irrelevant for the choice of optimal precision for the forecaster. The intuition for this is the following: since by assumption $x$ is independent of all other random variables (the state $\theta$ in particular), the forecaster's signal is uninformative about $x$. Hence, when policy uncertainty increases (keeping its mean fixed), the marginal benefit of acquiring information about the state does not change, and therefore the optimal amount of information is also unchanged.


\section{Extension: PM transparency}\label{sec: transparency}
An interesting question is whether PM transparency about her information will have a positive or negative impact on the incentives of the forecaster to gather private information. To look into this question, consider the case in which the PM releases her signal $s_P$ publicly. Then, the forecaster knows that the PM's action will be
\begin{align*}
a^*=-x E(\theta|s_P) = -x H s_P,
\end{align*} 
and furthermore, can condition his forecast on both $s$ and $s_P$. Plugging the PM's action into the forecaster's interim optimal forecast problem from equation (\ref{eq: forecaster's problem}), that problem becomes
\begin{align*}
\min_{f}E[(\theta+a^*-f)^2|s,s_P]=\min_{f}E[(\theta-x H s_P-f)^2|s,s_P].
\end{align*}

The form of the optimal forecast under PM transparency in the next lemma follows from standard results for quadratic loss functions, and the form of the optimal interim loss follows from direct calculation after substituting the optimal forecast into the interim loss.
\begin{lemma}\label{lemma: optimal forecast and interim loss transparency}
\begin{enumerate}
\item The optimal forecast with PM transparency is
\begin{align*}
f^*=E(\theta|s,s_P)- \mu H s_P.
\end{align*}

\item The optimal interim loss for the forecaster with PM transparency is
\begin{align}\label{eq: optimal interim loss transparency}
Var(\theta|s,s_P)+H^2 s_P^2 Var(x) + \tau^2.
\end{align}
\end{enumerate}
\end{lemma}

The last two terms in (\ref{eq: optimal interim loss transparency}) are independent of $k$, and therefore irrelevant for the choice of optimal precision. Furthermore, given the Gaussian information structure, the conditional variance $Var(\theta|s,s_P)$ is
\begin{eqnarray*}
Var(\theta|s,s_P)=\frac{1}{t+h+k}.
\end{eqnarray*}
Hence, the the ex-ante problem for the forecaster who observes the PM's signal is
\begin{eqnarray}\label{eq: forecaster's ex ante problem transparency}
\min_{k\ge 0}\left\{\frac{1}{t+h+k}+C(k)\right\}.
\end{eqnarray}
For comparison, in the case in which the PM did not release her signal the forecaster's problem was
\begin{eqnarray}\label{eq: forecaster's ex ante problem no transparency}
\min_{k\ge 0} \left\{\left(1-\frac{\mu h}{t+h}\right)^2 \left(\frac{1}{t+k}\right)+C(k)\right\}.
\end{eqnarray}

To compare problems (\ref{eq: forecaster's ex ante problem transparency}) and (\ref{eq: forecaster's ex ante problem no transparency}), it is useful to compute the marginal benefit of information in each problem. In the problem in which the PM releases her signal, the forecaster's marginal benefit of precision $k$ is
\begin{eqnarray}\label{eq: marginal benefit transparency}
\left(\frac{1}{t+h+k}\right)^2.
\end{eqnarray}
On the other hand, when the PM does not release her signal (the baseline problem), the forecaster's marginal benefit of precision $k$ is
\begin{eqnarray}\label{eq: marginal benefit no transparency}
\left(1-\frac{\mu h}{t+h}\right)^2 \left(\frac{1}{t+k}\right)^2.
\end{eqnarray}
Inspecting equations (\ref{eq: marginal benefit transparency}) and (\ref{eq: marginal benefit no transparency}), we see that PM's transparency can have an ambiguous effect on the marginal benefit of information, in which case transparency has also an ambiguous effect on the incentives for the forecaster to acquire information. In particular, while
\begin{eqnarray*}
\left(\frac{1}{t+h+k}\right)^2 < \left(\frac{1}{t+k}\right)^2,
\end{eqnarray*}
still we may have that the marginal benefit with transparency (\ref{eq: marginal benefit transparency}) may be higher than the marginal benefit with no transparency (\ref{eq: marginal benefit no transparency}) if $\left(1-\mu\frac{h}{t+h}\right)^2<1$.

However, if $\left(1-\mu\frac{h}{t+h}\right)^2>1$, we can unambiguously sign the effect of transparency on the marginal benefit of information. Proposition \ref{prop: transparency} below gives sufficient conditions under which PM transparency lowers the forecaster's marginal benefit of information, and hence crowds-out forecaster's information acquisition. We note that Proposition \ref{prop: transparency} gives only sufficient conditions for that to be the case, and that the crowding-out may even occur when the conditions in the proposition are not met.
\begin{proposition}\label{prop: transparency}
Assume that either $\mu\le0$ or $\mu\ge 2\frac{t+h}{h}$. Then PM's transparency about her signal decreases the marginal value of forecaster's precision, and the forecaster's optimal precision is weakly lower than when the PM does not release her signal.
\end{proposition}

To grasp the intuition of Proposition \ref{prop: transparency}, first note that when the PM discloses her information, the conditional variance of the state for the forecaster diminishes, then directly lowering the marginal benefit from information acquisition. Second, under the conditions of Proposition \ref{prop: transparency} the PM's action \emph{reinforces} the effect of the state on the outcome. Now, PM transparency makes that reinforcement irrelevant for the forecaster, since the PM's action is now perfectly predictable, and the forecaster can account for it in the forecast. On the other hand, without PM transparency that reinforcement increases the effect of the state on the outcome, hence increasing the marginal value of acquiring information. Therefore, under the conditions of Proposition \ref{prop: transparency} the marginal benefit of information is unambiguously higher with no transparency, and therefore transparency crowds-out information acquisition.

When the conditions of Proposition \ref{prop: transparency} are not met, it is possible that PM transparency may induce more information acquisition. The reason is that in that case PM's actions are \emph{preventive}, i.e. they attenuate the effect of the state on the outcome. Hence, while transparency does lower uncertainty about the state, it eliminates the attenuation perception for the forecaster -- recall that the forecaster now perfectly accounts for policy in the forecast. Therefore, in this case the effect of transparency on information acquisition is ambiguous.


\section{Conclusion}\label{sec: conclusion}
The model and analysis in this paper constitute an attempt to understand information acquisition incentives of forecasters in environments in which forecasted variables are impacted by policy interventions. While empirically relevant, this case has been mostly overlooked in the literature. Our goal is to bridge that gap.

Our results show that the policy-making environment can have a significant impact on the information acquisition incentives of forecasters. In particular, for plausible parameter values our analysis shows that: $(i)$ the forecaster acquires less private information than in the intervention-free case; $(ii)$ the optimal private precision of the forecaster decreases in the precision of PM's information; $(iii)$ the optimal private precision of the forecaster decreases in the strength of policy intervention. These results are consistent with prior empirical findings, which show that when forecasted variables subject to policy intervention, private forecasts of various forecasters are based on no proprietary information, and instead are made using the same publicly available data.

Moreover, in an extension of our model we look to how PM's transparency about her private information impacts the information acquisition incentives of forecasters. While for some parameter values the effect of transparency is ambiguous, we identify the region of the parameter space for which PM transparency does crowd-out forecaster's precision.

Taken together, our findings point to the important effects that intervention may have on forecasters' information acquisition incentives. In particular, we show that to properly understand the informational value of forecasts, we must pay close attention to the policy-making environment in which the forecasting exercise is embedded.

Finally, we point out that while in this paper we deal with a forecasting problem for concreteness, our results apply directly to any information acquisition problem in which: $(i)$ the payoff of the agent who must make the information acquisition decision (the forecaster in our paper) depends on a stochastic outcome and on the agent's own decision; and $(ii)$ the payoff-relevant outcome depends on an exogenous state and on the policy action of another agent (the PM in our paper). It is easy to think of other applications that fit the description in points $(i)$ and $(ii)$, as for example that of a bond investor mentioned in the introduction.

\bibliographystyle{econometrica}
\bibliography{/Users/gusti/Dropbox/Bibliography/references}

\section*{Appendix: Proofs}

{\footnotesize

\subsection*{Proof of Lemma \ref{lemma: optimal forecast and interim loss}}

The forecaster's interim problem is
\begin{eqnarray}\label{eq: interim forecaster's problem app}
\min_{f} E\{[(1-x H ) \theta-f]^2 | s\} + H^2 E(x^2) E(\eta^2)+\tau^2,
\end{eqnarray}

The last two terms in equation (\ref{eq: interim forecaster's problem app}) are independent of $f$, and hence irrelevant for the choice of an optimal forecast. Hence, the forecaster problem is that of minimizing the first term in (\ref{eq: interim forecaster's problem app}), and standard results for quadratic loss functions show that the optimal forecast is
\begin{eqnarray}\label{eq: optimal forecast app}
f^*= & E[(1-x H )\theta|s].
\end{eqnarray}

Substituting the optimal forecast (\ref{eq: optimal forecast app}) into the interim loss function (\ref{eq: interim forecaster's problem app}), we obtain the forecaster's optimal interim loss
\begin{eqnarray}\label{eq: optimal interim forecaster's loss app}
E \{[(1-x H )\theta-E((1-x H )\theta|s) ]^2|s\} + H^2 E(x^2) E(\eta^2) + \tau^2.
\end{eqnarray}
Notice that inside the expectation in the first term in equation (\ref{eq: optimal interim forecaster's loss app}) we have the square of the difference between the random variable 
\begin{eqnarray}\label{eq: auxiliary label app}
(1-x H)\theta
\end{eqnarray}
and its conditional mean
\begin{eqnarray*}
E[(1-x H)\theta|s)].
\end{eqnarray*}
Hence, the expectation in the first term of equation (\ref{eq: optimal interim forecaster's loss app}) is equal to the variance of (\ref{eq: auxiliary label app}) conditional on $s$. That is, we can write the optimal interim loss (\ref{eq: optimal interim forecaster's loss app}) as
\begin{eqnarray*}
& Var [(1-x H )\theta |s ]+ H^2 E(x^2) E(\eta^2) + \tau^2,
\end{eqnarray*}
as claimed.\eprf

\subsection*{Proof of Lemma \ref{lemma: ex-ante loss}}
The ex-ante problem of the forecaster is
\begin{eqnarray}\label{eq: forecaster's ex-ante problem app}
\min_{k\ge 0} E\{Var [(1-x H )\theta |s ]\}+C(k).
\end{eqnarray}
From the Law of Total Variance, the expectation in the first term in (\ref{eq: forecaster's ex-ante problem app}) is equal to
\begin{eqnarray}\label{eq: forecaster's ex-ante problem app 2}
E\{Var [(1-x H )\theta |s ]\}=Var [(1-x H )\theta ] - Var\{E[(1-x H)\theta |s]\}.
\end{eqnarray}
We can directly compute each of the terms in (\ref{eq: forecaster's ex-ante problem app 2}).

For the first term, given that $x$ and $\theta$ are independentwe have
\begin{eqnarray*}
Var [(1-x H )\theta ]=Var[(1-x H )] Var(\theta)+Var[(1-x H )] [E(\theta)]^2 +[E(1-x H )]^2 Var(\theta).
\end{eqnarray*}
Using the fact that the prior mean is zero, the above expression reduces to
\begin{align*}
Var [(1-x H )\theta ]=Var[(1-x H )] Var(\theta)+[E(1-x H )]^2 Var(\theta)= \frac{H^2 \sigma^2}{t} + \frac{(1-\mu H)^2}{t}.
\end{align*}
where we have written $\frac{1}{t}$ for the prior variance.

Turning to the second term in (\ref{eq: forecaster's ex-ante problem app 2}), given the Gaussian information structure and our independence assumptions, we have that
\begin{align*}
E[(1-x H)\theta |s] = & E[(1-x H)] E(\theta|s)\\
= & (1-\mu H) K s,
\end{align*}
where we have again used the fact that the prior mean is equal to zero, and we denote $K=\frac{k}{t+k}$. Parameter $K$ is the weight that the forecaster puts on his signal $s$ when estimating the state (the weight on the prior mean is $1-K$, and recall that the prior mean is normalized to zero). Plugging-in this conditional expectation, the second term in (\ref{eq: forecaster's ex-ante problem app 2}) is
\begin{align}
Var\{E[(1-x H)\theta |s]\} = & Var[(1-\mu H) K s] \nonumber\\
= & (1-\mu H)^2 K^2 Var(s)\nonumber\\
= & (1-\mu H)^2 K^2 Var(\theta+\delta)\nonumber\\
= & (1-\mu H)^2 \frac{k^2}{(t+k)^2} \left(\frac{1}{t}+\frac{1}{k}\right) \nonumber\\
= & \frac{1}{t} (1-\mu H)^2 \frac{k}{t+k}.  \nonumber
\end{align}

Putting both terms together, we have that expression (\ref{eq: forecaster's ex-ante problem app 2}) simplifies to
\begin{align*}
E\{Var [(1-x H )\theta |s ]\} = & Var [(1-x H )\theta ] - Var\{E[(1-x H)\theta |s]\}\\
= & \frac{H^2 \sigma^2}{t} + \frac{(1-\mu H)^2}{t} -  \frac{1}{t} (1-\mu H)^2 \frac{k}{t+k}\\
= & \frac{H^2 \sigma^2}{t} + \frac{(1-\mu H)^2}{t} \left[1-\frac{k}{t+k} \right]\\
= & \frac{H^2 \sigma^2}{t} + \frac{(1-\mu H)^2}{t} \left[\frac{t}{t+k} \right]\\
= & \frac{H^2 \sigma^2}{t} + (1-\mu H)^2 \left[\frac{1}{t+k} \right].
\end{align*}

Hence, the forecaster's ex-ante problem (\ref{eq: forecaster's ex-ante problem app}) is
\begin{eqnarray}\label{eq: forecaster's ex-ante problem app 3}
\min_{k\ge 0} E\{Var [(1-x H )\theta |s ]\}+C(k) = \min_{k\ge 0}  \frac{H^2 \sigma^2}{t} + (1-\mu H)^2 \left(\frac{1}{t+k} \right) + C(k).
\end{eqnarray}
Finally, since the first term in (\ref{eq: forecaster's ex-ante problem app 3}) is independent of the forecaster's precision $k$, the forecaster's ex-ante problem reduces to
\begin{eqnarray*}
\min_{k\ge 0}  (1-\mu H)^2 \left(\frac{1}{t+k} \right) + C(k),
\end{eqnarray*}
as claimed.\eprf

\subsection*{Proof of Proposition \ref{proposition: precisions comparison intervention no intervention}}
The forecaster's ex-ante problem with intervention is
\begin{eqnarray*}
\min_{k\ge 0}  \{(1-\mu H)^2 \left(\frac{1}{t+k} \right) + C(k)\},
\end{eqnarray*}
or equivalently,
\begin{eqnarray}\label{eq: forecaster's ex-ante problem intervention app}
& \max_{k\ge 0}  \{-(1-\mu H)^2 \left(\frac{1}{t+k} \right) - C(k)\} \nonumber\\
 & = \max_{k\ge 0}  \{-R \left(\frac{1}{t+k} \right) - C(k)\},
\end{eqnarray}
where $R=(1-\mu H)^2$. The objective function of problem (\ref{eq: forecaster's ex-ante problem intervention app}) is strictly concave (recall that $C(k)$ is assumed to be convex), and therefore the solution of (\ref{eq: forecaster's ex-ante problem intervention app}) is unique. Furthermore, the objective function has increasing differences in $(R,k)$. Therefore, standard monotone comparative statics results (see \cite{Milgrom:1994aa}, Theorem 5) imply that the optimizer $k^*$ of problem (\ref{eq: forecaster's ex-ante problem intervention app}) is weakly increasing in $R$. Since in the no-intervention case $R=1$, it follows that in the intervention case ($R\neq 1$) optimal precision is weakly higher than in the no-intervention case if $R>1$, and it is weakly lower than in the no-intervention case if $R<1$, as claimed.\eprf

\subsection*{Proof of Proposition \ref{prop: CS on marginal benefit}}
The forecaster's marginal benefit of precision is
\begin{align}\label{eq: marginal benefit app}
\frac{\partial g(k; h,t,\mu)}{\partial k}= \left(1-\mu \frac{h}{t+h}\right)^2  \left(\frac{1}{t+k}\right)^2>0.
\end{align}
\begin{enumerate}[(a)]
\item
The partial derivative of (\ref{eq: marginal benefit app}) with respect to $\mu$ is
\begin{eqnarray}\label{eq: marginal benefit derivative mu app}
\frac{\partial^2 g(k; h,t,\mu)}{\partial \mu \partial k} =-2  \left(1-\mu \frac{h}{t+h}\right) \left(\frac{h}{t+h}\right)  \left(\frac{1}{t+k}\right)^2>0.
\end{eqnarray}
The last two factors in (\ref{eq: marginal benefit derivative mu app}) are positive for $h>0$ and $t>0$. Hence, it is easily checked then that
\begin{eqnarray*}
\frac{\partial^2 g(k; h,t,\mu)}{\partial \mu \partial k}  \mathrel{\substack{>\\ = \\<}} \iff \left(1-\mu \frac{h}{t+h}\right) \mathrel{\substack{<\\ = \\>}} 0 \iff \mu \mathrel{\substack{>\\ = \\<}} \frac{t+h}{h}.
\end{eqnarray*}

\item
The partial derivative of (\ref{eq: marginal benefit app}) with respect to $h$ is
\begin{eqnarray}\label{eq: marginal benefit derivative h app}
\frac{\partial^2 g(k; h,t,\mu)}{\partial \mu \partial k} =-2 \mu  \left(1-\mu \frac{h}{t+h}\right) \frac{h}{(t+h)^2}  \left(\frac{1}{t+k}\right)^2>0.
\end{eqnarray}
The last two factors in (\ref{eq: marginal benefit derivative mu app}) are positive for $h>0$. Hence, it is straightforward to check that
\begin{eqnarray*}
\frac{\partial^2 g(k; h,t,\mu)}{\partial h \partial k}  \mathrel{\substack{>\\ = \\<}} \iff \mu \left(1-\mu \frac{h}{t+h}\right) \mathrel{\substack{ <\\ = \\>}} 0 \iff \mathrel{\substack{\mu  \in [-\bar{x},0) \cup (\frac{t+h}{h},\bar{x}]\\  \mu  \in \{0, \frac{t+h}{h}\} \\ \mu  \in (0,\frac{t+h}{h})}}.
\end{eqnarray*}

\item 
The marginal benefit (\ref{eq: marginal benefit app}) is positive, and its second factor $\left(\frac{1}{t+k}\right)^2$ is strictly decreasing in $t$. Hence, a sufficient condition for the marginal benefit to be strictly decreasing in $t$ is that its first factor $\left(1-\mu \frac{h}{t+h}\right)^2$ is decreasing in $t$, which is true if and only if $\mu  \in [-\bar{x},0] \cup [\frac{t+h}{h},\bar{x}]$.
\end{enumerate}

Putting together the results of parts (a)-(c) gives the comparative statics of the proposition.\eprf

\subsection*{Proof of Lemma \ref{lemma: optimal forecast and interim loss transparency}}
The forecaster's interim problem when observing the PM's signal is
\begin{align*}
\min_{f}E[(\theta-x H s_P-f)^2|s,s_P].
\end{align*}
Standard results for quadratic loss functions imply that the optimal forecast is
\begin{align*}
f^*=E(\theta|s,s_P)- \mu H s_P.
\end{align*}
Plugging the optimal forecast back into the forecaster's interim loss function, the optimal interim loss is
\begin{eqnarray*}
E\{[(\theta-E(\theta|s,s_P)- (x-\mu) H s_P]^2|s,s_P\}=Var(\theta|s,s_P)+H^2 s_P^2 Var(x),
\end{eqnarray*}
where we have used the fact that $x$ is independent of all other random variables.\eprf

\subsection*{Proof of Proposition \ref{prop: transparency}}
Consider the following function
\begin{eqnarray}
f(k,\alpha)=
\begin{cases}
-\frac{1}{t+h+k}-C(k) \qquad &  \alpha=0 \\
-R \frac{1}{t+k}-C(k) \qquad  & \alpha=1,
\end{cases}
\end{eqnarray}
where $R=\left(1- \mu\frac{ h}{t+h}\right)^2$. Then $f(k,0)$ is the negative of the expected loss of the forecaster with PM transparency, and $f(k,1)$ is the negative of the expected loss of the forecaster without PM transparency. Under the assumptions of the proposition $R\ge 1$. Then,
\begin{eqnarray}
\frac{\partial f(k,1)}{\partial k}=R \frac{1}{(t+k)^2} >  \frac{1}{(t+h+k)^2} =\frac{\partial f(k,0)}{\partial k}
\end{eqnarray}
for all $h>0$. That is, the function $\frac{\partial f(k,\alpha)}{\partial k}$ is strictly increasing in $\alpha$. This implies that $f(k,\alpha)$ has strictly increasing differences in $(k,\alpha)$; hence, standard monotone comparative statics results (see \cite{Milgrom:1994aa}, Theorem 5) imply that the optimal precision of the problem with $\alpha=1$, is at least as high as the optimal precision with $\alpha=0$. But the former is the optimal precision without PM information disclosure, while the latter is the optimal precision with PM's information disclosure. Hence, forecaster's optimal precision is weakly lower with PM transparency.\eprf

\footnotesize}

\end{document}